\documentclass[12pt]{article}

\usepackage{amssymb}
\usepackage{amsmath}
\usepackage{delarray}
\usepackage[T1]{fontenc}
\usepackage{bbold}
\usepackage{amsfonts}   

\newcommand{\Ah}{\ensuremath{\hat{\mathcal A}}}
\newcommand{\An}{\ensuremath{\mathcal A}}
\newcommand{\As}{\ensuremath{{\mathcal A}_\star}}
\newcommand{\C}{\ensuremath{\mathbb C}}
\newcommand{\Dal}{\ensuremath{D_{\alpha}}}

\newcommand{\Dmu}{\ensuremath{D_{\mu}}}
\newcommand{\Dnu}{\ensuremath{D_{\nu}}}
\newcommand{\ds}{\displaystyle}
\newcommand{\g}{\ensuremath{\mathfrak g}}

\newcommand{\Hh}{\ensuremath{\hat{\mathcal{H}}}}
\newcommand{\Hn}{\ensuremath{\mathcal{H}}}
\newcommand{\ka}{\ensuremath{\kappa}}

\newcommand{\lpl}{\ensuremath{\bigtriangleup}}

\newcommand{\Mmn}{\ensuremath{M_{\mu\nu}}}

\newcommand{\mD}{\ensuremath{{\mathcal D}}}
\newcommand{\mE}{\ensuremath{{\mathcal E}}}
\newcommand{\mK}{\ensuremath{{\mathcal K}}}
\newcommand{\mP}{\ensuremath{{\mathcal P}}}

\newcommand{\p}{\ensuremath{\mathfrak p}}
\newcommand{\pal}{\ensuremath{\partial_{\alpha}}}

\newcommand{\pmu}{\ensuremath{\partial_{\mu}}}
\newcommand{\pnu}{\ensuremath{\partial_{\nu}}}
\newcommand{\R}{\ensuremath{\mathbb R}}

\newcommand{\rhdb}{\ensuremath{\blacktriangleright}}

\newcommand{\xmu}{\ensuremath{x_{\mu}}}
\newcommand{\xnu}{\ensuremath{x_{\nu}}}

\newcommand{\cm}{\'{c}}

\newcommand{\ct}{\v{c}}

\begin{document}

\begin{center}
{\bf  \Large Kappa-Minkowski spacetime, kappa-Poincar\'{e} Hopf algebra and
  realizations}
 
\bigskip

D. Kova\ct evi\cm{\footnote{e-mail:domagoj.kovacevic@fer.hr}}\\
  Faculty of Electrical Engineering and Computing, Unska 3,
  HR-10000 Zagreb, Croatia\\[3mm]
S. Meljanac {\footnote{e-mail: meljanac@irb.hr}}\\  
  Ru\dj er Bo\v{s}kovi\'c Institute, Bijeni\v cka  c.54, HR-10002 Zagreb,
  Croatia \\[3mm]                      

\end{center}
\setcounter{page}{1}
\bigskip

\begin{abstract}
  We unify \ka-Minkowki spacetime and Lorentz algebra in unique Lie algebra.
  Introducing commutative momenta, a family of \ka-deformed Heisenberg
  algebras and \ka-deformed Poincare algebras are defined.
  They are specified by the matrix depending on momenta.
  We construct all such matrices.
  Realizations and star product are defined and analyzed in general and
  specially, their relation to coproduct of momenta is pointed out.
  Hopf algebra of the Poincare algebra, related to the covariant realization,
  is presented in unified covariant form.
  Left-right dual realizations and dual algebra are introduced and considered.
  The generalized involution and the {\itshape star} inner product are
  analyzed and their properties are discussed.
  Partial integration and deformed trace property are obtained in general.
  The translation invariance of the star product is pointed out.
  Finally, perturbative approach up to the first order in $a$
  is presented in Appendix.
\end{abstract}

\section{Introduction}

Currently there is a widespread belief that usual description of spacetime as
a continuum can no longer survive at the very short distances such
as those of the order of the Planck length \cite{dop1,dop2}. The physics at
such short distances thus might require a modification of
spacetime geometry. An appropriate setup for  working out the idea of this
kind is provided by the noncommutative geometry (NC) framework. The
indication for NC geometry, playing important role at Planck scale, comes from
combined application of general relativity and Heisenberg uncertainty
principle, which leads to a class of models  with spacetime noncommutativity
\cite{dop1,dop2}.

Independently of this consideration, the idea for introducing
non-trivial algebra of coordinates is not a recent one and has a
rather long history. First proposal for spacetime  noncommutativity
was made by Snyder \cite{snyder}, whose original motivation was to get rid of
divergences appearing in calculations of Feynman diagrams. The idea
that lied behind the scene was to use noncommutativity between space
coordinates as a mean for implementing a cut-off in quantum field
theory, resulting with regularization of divergences.
Additionally, since it was realized that the open string theories and
D-branes in the presence of a background antisymmetric B-field led to
effective noncommutative field theory \cite{Seiberg:1999vs}, subsequently,
there emerged the idea that field theories on noncommutative spaces can
capture certain generic features of quantum theory of gravity.

The noncommutativity of spacetime implies deformation of the algebra
of functions, so that smooth spacetime geometry of classical gravity
has to be replaced with a quantum group ( Hopf algebra) description
\cite{majid1} at the Planck scale. There are many models of spacetime
noncommutativity, which among others include $\kappa$-space, Moyal
space and Snyder space.
 Each of these three allows for the Hopf algebra
description and has its own physical motivation. For the Hopf algebra
description of $\kappa$-space see \cite{Lukierski:1991pn,Lukierski2,lr94,lrtn94,
Lukierski:1993wx,Majid:1994cy,zakrzewski}, for Moyal space see
\cite{S2,balachandran} and
corresponding coverage for Snyder space is given in
\cite{bks06,Battisti:2008xy,Battisti:2010sr,gl10,m11,ckt11,ls11}. It is also
possible to analyze algebraic structure for the NC spaces emerging as
various combinations of these three basic types of noncommutativity.
See for example \cite{hs06,mk08} for the Hopf algebra
description of spacetime noncommutativity combining between Moyal
space and $\kappa$-space and \cite{mmss1011} for interpolation
between $\kappa$-space and Snyder space.

The analysis carried out here deals exclusively with $\kappa$-type of
noncommutativity. It is believed that this type of spacetime
deformation arises as a low energy limit of quantum gravity coupled to
matter fields, where effective theory, obtained after integrating out
topological degrees of freedom of gravity, has a symmetry specified by
a $\kappa$-Poincar\'{e} group \cite{AmelinoCamelia:2003xp,Freidel:2003sp,
Freidel:2005bb,Freidel:2005me,Freidel:2005ec}.  
Another argument in favour of $\kappa$-Minkowski space is that it can
serve as a playground arena for developing and testing phenomenological
predictions coming from Doubly Special Relativity (DSR) theories
\cite{AmelinoCamelia:2000mn,AmelinoCamelia:2000ge,aca02,Magueijo:2001cr}.
Most of considerations on DSR has been made within the framework of
$\kappa$-Poincar\'{e} algebra, which is believed to describe a
symmetry lying behind these theories. One can add the coalgebra structure
to the $\kappa$-Poincar\'{e} algebra in order to obtain 
a Hopf algebra. The algebraic sector of this quantum algebra can be
given with different commutation relations between the generators of
the algebra, which corresponds to different basis of $\kappa$-Poincar\'{e}.
However, it is known that all these basis lead to the same type of
spacetime noncommutativity, set up by the $\kappa$-Minkowski spacetime
\cite{kgn,KowalskiGlikman:2002jr}.
The \ka-deformed Poincar\'{e} algebras and quantum Clifford-Hopf algebras
are studied in \cite{rbv}.
General quantum Poincar\'{e} groups are described and investigated in \cite{pw95}.

It has been known for some time that in order to study quantum field
theory in the ultra high energy regime, one has to reconsider the
notion of particle statistics, since such extreme conditions can cause
statistics to show certain exotic features.
These modified statistics can be naturally incorporated within the
body of physics by the use of deformation quantization of quantum groups
\cite{drinfeld,gghmm0809} where deformation is carried by the
appropriate twist operator \cite{gghmm0809,bklvy,klry08,bp2}.
Physically this leads to twisted statistics and the associated R-matrices, which
in this way appear in the context of quantum field theories in noncommutative
spacetimes \cite{yz0708,dlw0703,dlw0708,dlw0807,lw10}.

In this paper we are establishing numerous results valuable from both,
mathematical and physical point of view, generalizing previous results
on realizations \cite{mk08,mks,mst,mssg}.
Our motivation is to unify \ka-Minkowski spacetime and Lorentz algebra
in the unique Lie algebra \cite{mks}.
Introducing commutative momenta, a family of \ka-deformed Heisenberg
algebras and \ka-deformed Poincar\'{e} (Hopf) algebras are defined.
\ka-deformed Poincar\'{e} (Hopf) algebras are specified uniquely by the
matrix $[h_{\mu\nu}(p)]$ depending on momenta and we construct all of them.
We point out the notion of realization and its relation to the star product and
the coalgebra structure.
Also, important integral identities satisfied by the star product on \ka-space
are included, among which are quasicyclicity and deformed trace
properties, as well as Jacobians encompassing the true form of the
integral measure. The important issue is that all these results have
the smooth limit as the deformation parameter vanishes, which is
different from the previous discussion in the literature
\cite{Dimitrijevic:2003wv,Moller:2004sk,Agostini:2006zza}
concerning the same subject.
Furthermore, we put forward an alternative view on the issue of
translation invariance of the star product. In establishing these results we took
over the methods from
\cite{fockspace,bardek} in connection with Fock space analysis.

The plan of the paper is as follows. In the Section \ref{ksp}, we consider
\ka-Minkowski spacetime and \ka-Poincar\'{e} algebra.
The consistency relations obtained from Jacobi identities are constructed and
analyzed. Also, the action \rhdb\ is introduced as a
preparation for the construction of the Hopf algebra structure.
In the Section \ref{rsp}, we mention the notion of
the realization,  the action $\rhd$ and the star product.
After the example of the {\itshape natural} realization, the structure of the
\ka-Poincar\'e Hopf algebra is obtained. The relation among the star
product, coproduct and realization is analyzed.
In the Section \ref{crk}, it is shown how to obtain any
realization from one particular (e.g. {\itshape natural}) realization. This
construction also produces the coalgebra structure. The construction is
followed by several examples.
Important observation is duality of realizations, which is explained in the
Section \ref{lr}. The {\itshape generalized} involution and the {\itshape star}
inner product are introduced in the Section \ref{if}. Also, several properties of
{\itshape star} inner product are analyzed. The translation invariance is
defined and explained in the Section \ref{ti}.
Main results are repeated in Section \ref{con}. Also, some ideas about future
work are given.
Finally, in the Appendix, linear approximation in $a$ is calculated as a nice
example, useful for applications.

\section{\ka-Minkowski spacetime and \ka-Poincar\'e algebra}
\label{ksp}

Let $\hat{x}_0,\hat{x}_1,\ldots,\hat{x}_{n-1}$ be the coordinates of the
\ka-Minkowski space. The commutation relations are given by
\begin{equation}
  [\hat{x}_\mu,\hat{x}_\nu]=i\left(a_\mu\hat{x}_\nu-a_\nu\hat{x}_\mu\right)
  \label{b1}
\end{equation}
for some vector $a=(a_0,\ldots,a_{n-1})\in\R^n$ (see \cite{l01,klm00}).
Our \ka-Minkowski space is a solvable Lie algebra with structure constants
$i(a_\mu\eta_{\nu\lambda}-a_\nu\eta_{\mu\lambda})$ and it will be
denoted by $\mathfrak{m}_\kappa$. Latin indices will be used for the
set $\{1,\ldots,n-1\}$ and Greek indices will be used for the set $\{0,\ldots,
n-1\}$. We denote the metric of our Minkowski space by $[\eta_{\mu\nu}]=
diag(-1,1,\ldots,1)$. The majority of results will be derived for any
vector $a$. However, in some situations, we will restrict our attention to
the vector $a$ for which $a_i=0$. It will simplify our calculations.

Let $\mathfrak{l}$ be the Lorentz algebra generated by \Mmn\ satisfying the
usual commutation relations
\begin{equation}
  [\Mmn,M_{\lambda\rho}]=M_{\mu\rho}\eta_{\nu\lambda}-M_{\nu\rho}\eta_{\mu
  \lambda}-M_{\mu\lambda}\eta_{\nu\rho}+M_{\nu\lambda}\eta_{\mu\rho}.
  \label{b4}
\end{equation}
Our \ka-Minkowski space $\mathfrak{m}_\ka$ can be enlarged to the Lie
algebra $\g_\ka$ if the following commutation relations
\begin{equation}
  [M_{\mu\nu},\hat{x}_\lambda]=\hat{x}_\mu\eta_{\nu\lambda}-\hat{x}_\nu
    \eta_{\mu\lambda}-i\left(a_\mu M_{\nu\lambda}-a_\nu M_{\mu\lambda}\right)
  \label{b7}
\end{equation}
are satisfied. One can check that all Jacobi identities are satisfied. Let us
mention that relation (\ref{b7}) is the only nontrivial relation such that
$\mathfrak{m}_\kappa$ together with $\mathfrak{l}$ form the Lie algebra
(see \cite{mks}).

The momentum generators $p_0,\ldots,p_{n-1}$ satisfy the
usual commutation relations
\begin{equation}
  [p_\mu,p_\nu]=0.
  \label{b10}
\end{equation}

We want to analyze the algebra \Hh\ generated by \Mmn, $\hat{x}_\mu$
and momenta $p_\mu$ with respect to some commutation relations. Before that,
let us consider remaining two commutation relations:
\begin{equation}
  [\hat{x}_\nu,p_\mu]=ih_{\mu\nu}(p)
  \label{b13}
\end{equation}
and
\begin{equation}
  [M_{\mu\nu},p_\lambda]=g_{\mu\nu\lambda}(p).
  \label{b16}
\end{equation}
Functions $h_{\mu\nu}$ and $g_{\mu\nu\lambda}$ are real, satisfying the
limit conditions
\begin{equation}
  \lim_{a\rightarrow0}h_{\mu\nu}=\eta_{\mu\nu}\cdot1,
  \label{b17}
\end{equation}
\begin{equation}  \det h\neq0
  \label{b18}
\end{equation}
and
\begin{equation}
  \lim_{a\rightarrow0}g_{\mu\nu\lambda}=
  p_\mu\eta_{\nu\lambda}-p_\nu\eta_{\mu\lambda}.
\end{equation}
We ask that all generators satisfy Jacobi identities.
Then, \Mmn, $\hat{x}_\mu$ and $p_\mu$ generate the algebra \Hh,
see also \cite{bp11}.
Jacobi identity is satisfied if all three generators are in $\mathfrak{g}_\ka$ or
two or three generators are in the algebra generated by $p_\mu$.
It remains to consider situations when one generator is $p_\mu$.

Jacobi identity for $M_{\mu\nu}$, $M_{\lambda\rho}$ and $p_\sigma$ produces
the differential equation
\begin{equation}
  \frac{\partial g_{\lambda\rho\sigma}}{\partial p_\alpha}g_{\mu\nu\alpha}-
  \frac{\partial g_{\mu\nu\sigma}}{\partial p_\alpha}g_{\lambda\rho\alpha}=
  g_{\mu\rho\sigma}\eta_{\nu\lambda}-g_{\nu\rho\sigma}\eta_{\mu\lambda}-
  g_{\mu\lambda\sigma}\eta_{\nu\rho}+g_{\nu\lambda\sigma}\eta_{\mu\rho}.
  \label{b20}
\end{equation}
Jacobi identity for $\hat{x}_\mu$, $\hat{x}_\nu$ and $p_\lambda$ produces the
differential equation
\begin{equation}
  \frac{\partial h_{\lambda\nu}}{\partial p_\alpha}h_{\alpha\mu}-
  \frac{\partial h_{\lambda\mu}}{\partial p_\alpha}h_{\alpha\nu}=
  a_\mu h_{\lambda\nu}-a_\nu h_{\lambda\mu}.
  \label{b22}
\end{equation}
Finally, the Jacobi identity for \Mmn, $p_\lambda$ and $\hat{x}_\lambda$
produces the differential equation
\begin{eqnarray}
  \frac{\partial g_{\mu\nu\lambda}}{\partial p_\alpha}h_{\alpha\rho}-
  \frac{\partial h_{\lambda\rho}}{\partial p_\alpha}g_{\mu\nu\alpha}=
  h_{\lambda\nu}\eta_{\mu\rho}-h_{\lambda\mu}\eta_{\nu\rho}+
  a_\mu g_{\nu\rho\lambda}-a_\nu g_{\mu\rho\lambda}.
  \label{b25}
\end{eqnarray}
See also \cite{mk0911}.

The next step is to analyze solutions of system of differential equations
(\ref{b20}), (\ref{b22}) and (\ref{b25}).
It is important to mention that functions $g_{\mu\nu\lambda}(p)$ are completely
determined by functions $h_{\mu\nu}(p)$ (see Section \ref{crk} for details).
At the same time, for one choice of functions $g_{\mu\nu\lambda}(p)$, it is
possible to construct an infinite family of functions $h_{\mu\nu}(p)$, as it
will be shown below.

In \cite{mks}, it is shown how to obtain a large family of solutions.
In Section \ref{crk} it will  be shown how to obtain the general solution of
the above system from one particular solution. In order to obtain one
particular solution, let us set $g_{\mu\nu\lambda}=p_\mu\eta_{\nu\lambda}-
p_\nu\eta_{\mu\lambda}$ i.~e. let us consider the commutator
\begin{equation}
  [\Mmn,p_\lambda]=p_\mu\eta_{\nu\lambda}-p_\nu\eta_{\mu\lambda}
  \label{b27},
\end{equation}
instead of commutator (\ref{b16}).
We note that $p_\lambda$ transforms as a vector under the action of \Mmn.
The solution of this system has the form
\begin{equation}
  h_{\mu\nu}=\eta_{\mu\nu}(-A+f(B))-ia_\mu p_\nu+a^2p_\mu p_\nu
  \gamma_2(B),
  \label{b28}
\end{equation}
where $A=-(ap)$, $B=-a^2p^2$,
\begin{equation}
  \gamma_2=-\frac{1+2f(B)\frac{df(B)}{dB}}{f(B)-2B\frac{df(B)}{dB}}
  \label{b29}
\end{equation}
and $f(B)$ is a real function. The scalar product $a_\alpha p^\alpha$ is
denoted by $(ap)$.

The simplest possible expression for $h_{\mu\nu}$ is obtained when
$\gamma_2=0$. Then relation (\ref{b29}) shows that $f(B)=\sqrt{1-B}$.
This case is related to {\itshape classical basis}
(see \cite{kgn,KowalskiGlikman:2002jr,bp10}) or {\itshape natural} realization
(see \cite{mks} or Subsection \ref{nr}). In order to distinguish this case from
other cases, we write $P_\mu$ instead of $p_\mu$. Similarly, we will use
capital letters for the {\itshape natural} realization.

If the commutator (\ref{b27}) is satisfied, then \Mmn\ and $p_\mu$ form
the Poincar\'{e} algebra $\mathfrak{so}(1,n-1)$. If more general
commutator (\ref{b16}) is satisfied then \Mmn\ and $p_\mu$
form \ka-Poincar\'{e} algebra. These algebras will be denoted by \p.

\subsection{The shift and Casimir operators}
\label{sco}

It is useful to consider the shift operator $Z$ and the Casimir operator
$\Box$. The invertible operator $Z$ is defined by
\begin{equation}
  [Z,\hat{x}_\mu]=ia_\mu Z
\end{equation}
or, equivalently
\begin{equation}
  [Z^{-1},\hat{x}_\mu]=-ia_\mu Z^{-1}
  \label{b31}
\end{equation}
and
\begin{equation}
  [Z,p_\mu]=0.
\end{equation}
Our $Z$ is called a shift operator since it satisfies relations
\begin{equation}
  Z\hat{x}_\mu Z^{-1}=\hat{x}_\mu+ia_\mu,\;\;\;
  Z^{-1}\hat{x}_\mu Z=\hat{x}_\mu-ia_\mu.
  \label{b32}
\end{equation}
Let us mention that $Z=Z(p)$.
If the commutator (\ref{b27}) is satisfied ($p_\mu$ transforms as a vector under
\Mmn), then $Z^{-1}$ has the form
\begin{equation}
  Z^{-1}=\frac{-A+f(B)}{\sqrt{(f(B))^2+B}}.
  \label{b34}
\end{equation}

The operator $\Box$\footnote{It would be more precise to write $\Box_h$ instead
of $\Box$ because it depends on the set of functions $h_{\mu\nu}$,
but we will omit the index in order to simplify the notation.}
is defined by commutation relations
\begin{equation}
  [\Mmn,\Box]=0,
\end{equation}
\begin{equation}
  [p_\mu,\Box]=0
\end{equation}
and
\begin{equation}
  [\Box,\hat{x}_\mu]=2ip_\mu.
\end{equation}
If the commutator (\ref{b27}) is satisfied, it is possible to show that $\Box$
has the form $\Box=\frac1{a^2}F(B)$ and then it is easy to get
\begin{equation}
  \Box=\int_0^B\frac{dt}{f(t)-t\gamma_2}.
\end{equation}
See also \cite{mmss1011}.

\subsection{The algebra \Hh\ and the action \rhdb}

Let us define the algebra \Hh\ as the quotient of the free algebra generated by
$1$, $\hat{x}_\mu$ and $p_\mu$ and the ideal generated by relations
(\ref{b1}), (\ref{b10}) and (\ref{b13}). Let us recall that all Jacobi
identities are satisfied.

We will show in Section \ref{crk} that Lorentz generators \Mmn\ are in the
algebra \Hh. Hence, the Poincar\'{e} algebra \p, generated
by \Mmn\ and $p_\mu$ is in \Hh. Let $\mathcal{P}$ be the enveloping
algebra of \p\ and let \Ah\ be the subspace
(subalgebra) of \Hh\ generated by $1$ and $\hat{x}_\mu$. Our goal
is to construct the action of \Hh\ on \Ah\ and then reconstruct the coalgebra
structure of the Poincar\'{e} algebra \p. Let us define the action \rhdb\ of
\Hh\ on \Ah\ by
\begin{enumerate}
  \item $\hat{x}_\mu\rhdb\hat{g}(\hat{x})=\hat{x}_\mu\hat{g}(\hat{x}),\;\;
    \hat{g}(\hat{x})\in\Ah$
  \item $p_\mu\rhdb1=0,\;\Mmn\rhdb1=0$
  \item $p_\mu\rhdb\hat{g}(\hat{x})=[p_\mu,\hat{g}(\hat{x})]\rhdb1=
    p_\mu\hat{g}(\hat{x})\rhdb1$\\
    $\Mmn\rhdb\hat{g}(\hat{x})=[\Mmn,\hat{g}(\hat{x})]\rhdb1=
    \Mmn\hat{g}(\hat{x})\rhdb1$.
\end{enumerate}
The action \rhdb\ is well defined since all Jacobi identities are satisfied.
Hence, \Ah\ is an \Hh-module.

\subsection{Leibniz rule and bialgebra}

In order to obtain the Hopf algebra structure from our Poincar\'e algebra,
we have to define coproduct \lpl, antipode $S$ and counit $\epsilon$ of
our generators. We start with the action of $p_\mu$.
\begin{eqnarray}
  p_\mu\rhdb(\hat{f}(\hat{x})\cdot\hat{g}(\hat{x}))&=&
    p_\mu\rhdb(\hat{f}(\hat{x})\rhdb\hat{g}(\hat{x}))=
    (p_\mu\hat{f}(\hat{x}))\rhdb\hat{g}(\hat{x})=\nonumber \\
  &=&\left([p_\mu,\hat{f}(\hat{x})]\right)\rhdb\hat{g}(\hat{x})+
    \left(\hat{f}(\hat{x})p_\mu\right)\rhdb\hat{g}(\hat{x})= \\
  &=&\left([p_\mu,\hat{f}(\hat{x})]\right)\rhdb\hat{g}(\hat{x})+
    m\left(\left(1\otimes p_\mu\right)\rhdb\left(\hat{f}(\hat{x})\otimes\hat{g}
    (\hat{x})\right)\right)\nonumber
\end{eqnarray}
where $m(a\otimes b)=a\cdot b$ and $[p_\mu,\hat{f}(\hat{x})]$ is written in
the form that the coproduct can be recognized. Since
\begin{equation}
  p_\mu\rhdb\left(\hat{f}(\hat{x})\cdot \hat{g}(\hat{x})\right)=
  m\left(\lpl p_\mu(\hat{f}(\hat{x})\otimes\hat{g}(\hat{x}))\right)
  \label{b37}
\end{equation}
(Leibniz rule, \cite{msh}), we have
\begin{equation}
  m\left(\lpl p_\mu\rhdb(\hat{f}(\hat{x})\otimes\hat{g}(\hat{x}))\right)=
  [p_\mu,\hat{f}(\hat{x})]\hat{g}(\hat{x})+
  m\left((1\otimes p_\mu)\rhdb(\hat{f}(\hat{x})\otimes\hat{g}(\hat{x}))\right)
  \label{b38}
\end{equation}
It shows that the coproduct $\lpl p_\mu$ can be obtained from formulae
$[p_\mu,\hat{f}(\hat{x})]$ by adding $1\otimes p_\mu$. The same calculation
can be done also for \Mmn.

It is important to emphasize that our construction produces necessary conditions
for the coproduct. However, for each particular case, all axioms for the
bialgebra have to be checked.

Let us point out an important formula generalizing equation (\ref{b1}).
Relation (\ref{b1}) shows that
\begin{equation}
  \hat{x}_\mu\hat{x}_{\nu}=ia_\mu\hat{x}_\nu+(\hat{x}_\nu-ia_\nu)
    \hat{x}_\mu
\end{equation}
and by using relation (\ref{b32}) it follows that
\begin{equation}
  \hat{x}_\mu\hat{x}_{\nu}=ia_\mu\hat{x}_\nu+Z^{-1}\hat{x}_\nu Z
    \hat{x}_\mu.
  \label{b40}
\end{equation}
Using the induction, one can obtain that
\begin{eqnarray}
  \hat{x}_\mu\hat{x}_{\nu_1}\ldots\hat{x}_{\nu_k}=
  Z^{-1}\hat{x}_{\nu_1}\ldots\hat{x}_{\nu_k}Z\hat{x}_\mu+ia_\mu
  \sum_{s=0}^{k-1}Z^{-1}\hat{x}_{\nu_1}\ldots\hat{x}_{\nu_s}Z
  \hat{x}_{\nu_{s+1}}\ldots\hat{x}_{\nu_k}.
  \label{b43}
\end{eqnarray}
By induction, again, the last expression can be written in the form
\begin{eqnarray}
  \hat{x}_\mu\hat{x}_{\nu_1}\ldots\hat{x}_{\nu_k}=
  Z^{-1}\hat{x}_{\nu_1}\ldots\hat{x}_{\nu_k}Z\hat{x}_\mu-
  a_\mu \left(p_\alpha^L\rhdb\hat{x}_{\nu_1}\ldots\hat{x}_{\nu_k}\right)
  \hat{x}^\alpha,
  \label{b46}
\end{eqnarray}
where $p_\alpha^L$ is the abbreviation for $p_\alpha^L=P_\alpha-
\frac{a_\alpha}2\Box$.
Important property of $p_\alpha^L$ is
\begin{equation}
  [p_\alpha^L,\hat{x}_\mu]=\eta_{\alpha\mu}Z^{-1}.
  \label{b49}
\end{equation}
Using the formula (\ref{b46}), it is easy to write down the Leibniz rule for
$\hat{x}_\mu$,
\begin{equation}
  \hat{x}_\mu\rhdb(\hat{f}\hat{g})=(Z^{-1}\rhdb\hat{f})(\hat{x}_\mu\rhdb
    \hat{g})-a_\mu(p_\alpha^L\rhdb\hat{f})(\hat{x}_\alpha\rhdb\hat{g})
  \label{b52}
\end{equation}

\section{Realizations and star product}
\label{rsp}

\subsection{Realizations and the action $\rhd$}

We are interested in realizations of our algebra \Hh\ in terms of the Weyl
algebras \Hn\ \cite{mks,mst,dmsz,fkgn06,mssg}.
Hence, $p_\mu$ and \Mmn\ will have the form $p_\mu(x,\partial)$
and $\Mmn(x,\partial)$.
In particular, we are interested in realizations of the subalgebra \p, defined
before the Subsection \ref{sco}. By abuse of notation, we will denote the
corresponding subalgebra of \Hn\ by \p\ again.
The Weyl algebra \Hn\ is the quotient of the free algebra generated by
$1$, \xmu\ and $\ds \pmu=\frac{\partial}{\partial x^\mu}$ and the
ideal generated by relations
\begin{equation}
  [\pmu,\pnu]=0,\;\;[\xmu,\xnu]=0,\;\;[\pmu,\xnu]=\eta_{\mu\nu}.
  \label{c1}
\end{equation}
Generators $\hat{x}_\mu$ and $p_\mu$ are expressed by \xmu\ and \pmu:
\begin{equation}
  p_\mu=-i\pmu,
  \label{c2}
\end{equation}
\begin{equation}
  \hat{x}_\mu=x^\alpha\varphi_{\alpha\mu}(\partial)
  \label{c4}
\end{equation}
The relationship between $\varphi$ and $h$ is given by
\begin{equation}
  \varphi_{\alpha\mu}(\partial)=h_{\alpha\mu}(p)=h_{\alpha\mu}(-i\partial).
  \label{c5}
\end{equation}

The relations (\ref{c2}), (\ref{c4}) and (\ref{c5}) show that \Hh\ is a subalgebra
of \Hn. At the same time relation (\ref{b17}) shows that \Hh\ is isomorphic to \Hn.

Our realization (\ref{c4}) is generally nonhermitian realization because
$\hat{x}_\mu$ is generally not a Hermitian operator. The advantage of
this realization is the polarization i.~e. $x$s are on the left hand side and
$\partial$s are on the right hand side. We will consider Hermitian realizations
in the next paper.

Let us consider the action $\rhd$ of \Hn\ on the subalgebra \An\ generated by
$1$ and \xmu. The action is given by
\begin{enumerate}
  \item $\xmu\rhd g(x)=\xmu g(x),\;\;g(x)\in\An$
  \item $\pmu\rhd g(x)=[\pmu,g(x)]\rhd1=\pmu g(x)\rhd1$
\end{enumerate}

If the relation (\ref{b27}) is satisfied,
it follows from (\ref{b28}) that $\hat{x}_\mu$ can be written in the form
\begin{equation}
  \hat{x}_\mu=x_\mu(-A+f(B))+i(ax)\partial_\mu-a^2(x\partial)\partial_\mu
    \gamma_2
  \label{c6}
\end{equation}
where $A=i(a\partial)$ and $B=a^2\partial^2$.

Let us consider one example. We will calculate the commutator
$[\hat{x}_\mu,f(\hat{x})]$ using the approximation of $\hat{x}$ in the first
order in $a$ (see also the Appendix).
Let us mention that it can be also considered as a generalization of
formula (\ref{b1}), since the right hand side of (\ref{b1}) is linear in $a$.
If $f(\hat{x})$ has the form of the monomial, then
\begin{eqnarray}
  [\hat{x}_\mu,\hat{x}_{k_1}\ldots\hat{x}_{k_m}]&=&
    \sum_{l=1}^m\hat{x}_{k_1}\ldots\hat{x}_{k_{l-1}}[\hat{x}_\mu,\hat{x}_{k_l}]
    \hat{x}_{k_{l+1}}\ldots\hat{x}_{k_m}=\nonumber \\
  &=&\sum_{l=1}^m\hat{x}_{k_1}\ldots\hat{x}_{k_{l-1}}
    i(a_\mu\hat{x}_{k_l}-a_{k_l}
    \hat{x}_\mu)\hat{x}_{k_{l+1}}\ldots\hat{x}_{k_m}=\nonumber \\
  &=&ima_\mu \hat{x}_{k_1}\ldots\hat{x}_{k_m}
    -\sum_{l=1}^m\hat{x}_{k_1}\ldots\hat{x}_{k_{l-1}}ia_{k_l}
    \hat{x}_\mu\hat{x}_{k_{l+1}}\ldots\hat{x}_{k_m}.\nonumber
\end{eqnarray}
Since all terms already contain $a$, $\hat{x}_\mu$ can be replaced by
\xmu\ and above expression transforms to
\begin{equation}
  i\left(a_\mu(x\partial)-\xmu(a\partial)\right)\left(x_1\ldots x_{k_m}\right)
\end{equation}
It is interesting to note that this expression (in the first order in $a$)
does not depend on the realization $\varphi$.

\subsection{Star product}

For the given realization $\varphi$ (relation (\ref{c4})) it is possible to
construct the bijection $T$ from \Ah\ to \An\ by
\begin{equation}
  T(\hat{x}_\mu)=\hat{x}_\mu\rhd1=x^\alpha\varphi_{\alpha\mu}
  \rhd1=\xmu
\end{equation}
and
\begin{equation}
  T(\hat{g}(\hat{x}))=\hat{g}(\hat{x}_\mu)\rhd1=g(x).
  \label{c41}
\end{equation}
Let us mention that the polarization of the realization (\ref{c4}) is
important for our bijection.

Now, let us define the star product of functions $f$ and $g$ by
\begin{equation}
  (f\star g)(x)=\hat{f}(\hat{x})\hat{g}(\hat{x})\rhd1=
  \hat{f}(\hat{x})\rhd g(x)
  \label{c43}
\end{equation}
The star product depends on the realization $\varphi$.
Also, the star product is associative if the \ka-Minkowski space is of the Lie
type (Eq. (\ref{b1})).

The algebra \Ah\ is not isomorphic to algebra \An\ since the multiplication in
\Ah\ is not commutative. Hence, let \As\ be the algebra which has the same
elements as \An, but the star product is used instead of pointwise multiplication.
Then the definition of the star product (relation \ref{c41})
shows that $T:\Ah\rightarrow\As$ is an isomorphism of algebras.
Also, this isomorphism can be enlarged to the space of the formal power series.
Since $g(x)\rhdb1=\hat{g}(\hat{x})$ and $\hat{g}(\hat{x})\rhd1=g(x)$,
for $g(x)\in\An$, the action $\rhd$ is in some way the inverse action of the
action \rhdb.

Let us consider the {\itshape left covariant} realization. Then
$T(1)=1$, $T(\hat{x}_\mu)=x_\mu$ and
$T(\hat{x}_\mu\hat{x}_\nu)=x_\mu(x_\nu-ia_\nu)$.
Similarly, for the {\itshape right covariant} realization,
$T(\hat{x}_\mu\hat{x}_\nu)=(x_\mu+ia_\mu)x_\nu$.

The star product can be also written in the form
\begin{equation}
  (f\star g)(x)=m\left[:e^{x_\alpha(\lpl-\lpl_0)\partial^\alpha}:f\otimes g\right],
  \label{c46}
\end{equation}
where $:e^{x_\alpha\partial^\alpha}:$ denotes the polarization i.~e.
\begin{equation}
  :e^{x_\alpha \partial^\alpha}:=\sum_{k=0}^\infty\frac1{k!}x^k\partial^k=
  \sum_{k=0}^\infty\frac1{k!}\prod x_0^{\alpha_0}\cdots x_{n-1}^{\alpha_{n-1}}
  (\partial^0)^{\beta_0}\cdots(\partial^{n-1})^{\beta_{n-1}}
\end{equation}
(see \cite{mks, msh,mst,mssg} for details.)

Let us consider the function $\hat{f}(\hat{x})=e^{ik\hat{x}}$. Then, we define
functions $\mK_\varphi$ and $\mD_\varphi$ by
\begin{equation}
  e^{ik\hat{x}}\rhd1=e^{i(\mK_\varphi)_\alpha(k)x^\alpha}=e^{i\mK_\varphi(k) x}.
  \label{c49}
\end{equation}
and
\begin{equation}
  e^{ikx}\star e^{iqx}=e^{i\mD_\varphi(k,q)x}
  \label{c52}
\end{equation}
For the {\itshape totally symmetric} realization (or {\itshape Weyl-symmetric}
realization, see Section \ref{crk}), $\mK_s=id$ (the identity operator) and
then $\mD_s$ can be calculated using the BCH formula.

It can be proved that
\begin{equation}
  \lpl\pmu=i\mD_\varphi\left((-i\partial)\otimes1,1\otimes(-i\partial)\right),
  \label{c58}
\end{equation}
i.~e. $\mD_\varphi$ can be treated as a coproduct. Finally, it is easy to find
relation between $\mD_\varphi$ and $\mD_s$:
\begin{equation}
  \mD_\varphi(k,q)=\mK_\varphi(\mD_s(\mK_\varphi^{-1}(k),\mK_\varphi^{-1}(q))).
  \label{c61}
\end{equation}
Details can be found in \cite{mks,mssg,msa}.

In order to simplify the notation, we will omit the index $\varphi$ and write
\mK\ and \mD\ instead of $\mK_\varphi$ and $\mD_\varphi$.

It is useful to define the function \mP\ by
$e^{ik\hat{x}}\rhd e^{iqx}=e^{i\mP(k,q)x}$ (see also \cite{mss}). Then
$\mD(k,q)=\mP(\mK^{-1}(k),q)$.

Let us mention the relationship between the antipode and functions \mK\ and \mD.
The antipode and the function \mK\ commute,
\begin{equation}
  S(\mK(k))=\mK(S(k)).
  \label{c70}
\end{equation}
Also,
\begin{equation}
  \mK^{-1}(S(k))=-\mK^{-1}(k)
  \label{c71}
\end{equation}
(see \cite{msa}).
Similar formula is valid for the antipode and the function \mD\, but one should
note the order of variables,
\begin{equation}
  S(\mD(k,q))=\mD(S(q),S(k)).
  \label{c73}
\end{equation}
Formulae (\ref{c70}) and (\ref{c73}) will be used in section \ref{if}.

Our ordering is totally symmetric. However, to any realization $\chi$, it is
possible to associate an ordering $:\cdot:_\chi$ by the formula
\begin{equation}
  :e^{ik\hat{x}_\varphi}:_\chi\rhd1=e^{i\mK_\varphi^\chi(k)x}
\end{equation}
and the relation $\mK_\varphi^\varphi=id$. In our case, $\chi=S$, where
$S$ stands for the totally symmetric ordering. See more details in \cite{mssg}.

Finally, let us mention that $\ds \pmu(f\star g)=m_*\lpl\pmu(f\otimes g)$
where $m_*$ is the star multiplication i.~e. $m_*(f\otimes g)=f\star g$.

Also, the Leibniz rule for $\hat{x}_\mu$, \ref{b52}, can be written in terms of
realizations and the action $\rhd$.

\subsection{The natural realization}
\label{nr}

If the function $f(B)$ (equation (\ref{c6}) has the form $f(B)=\sqrt{1-B}$,
then this particular realization is called the {\itshape natural} realization
(see \cite{mks,msa}) or  the {\itshape classical basis} (see \cite{kgn,bp10}).
It is easy to see that
$\gamma_2=0$ (\ref{b29})). For the {\itshape natural} realization,
the Dirac derivative $D_\mu$ is equal to \pmu. In order to distinguish this
realization from other realizations, we will write $X_\mu$ and \Dmu\ instead
of \xmu\  and \pmu.

In our {\itshape natural} realization, $Z^{-1}$ and $\Box$ have the form
\begin{equation}
  Z^{-1}=-i(aD)+\sqrt{1-a^2D^2}
\end{equation}
and
\begin{equation}
  \Box=\frac2{a^2}\left(1-\sqrt{1-a^2D^2}\right).
  \label{c9}
\end{equation}
Our $\hat{x}_\mu$ has the form
\begin{equation}
  \hat{x}_\mu=X_\mu Z^{-1}+i(aX)\Dmu
\end{equation}
and formula (\ref{c4}) leads to
\begin{equation}
  [\Dmu,\hat{x}_\nu]=Z^{-1}\eta_{\mu\nu}+ia_\mu\Dnu.
  \label{c10}
\end{equation}

In order to develop a coalgebra structure on our Poincar\'e algebra, several
formulae have to be derived. The first step is to find $[\Dmu,\hat{f}
(\hat{x})]$. Using the formula (\ref{c10}) one obtains
\begin{eqnarray}
  [\Dmu,\hat{f}(\hat{x})]&=&\left(\Dmu\rhdb\hat{f}(\hat{x})\right)Z^{-1}+
  ia_\mu\left(\Dal Z\rhdb\hat{f}(\hat{x})\right)D^\alpha-\nonumber\\
  &-&\frac{ia_\mu}2\left(\Box Z\rhdb\hat{f}(\hat{x})\right)ia_\alpha D^\alpha.
\end{eqnarray}
In the Section \ref{crk}, the {\itshape left covariant} realization is introduced
and the relationship between the {\itshape left covariant} realization and the
{\itshape natural} realization is given by the formula (\ref{d34}). Hence,
the expression $\Dal-\frac{ia_\alpha}2\Box$ will be denoted by $\pal^L$ and
the previous formula transforms to
\begin{equation}
  [\Dmu,\hat{f}(\hat{x})]=\left(\Dmu\rhdb\hat{f}(\hat{x})\right)Z^{-1}+
  ia_\mu\left(\pal^L Z\rhdb\hat{f}(\hat{x})\right)D^\alpha.
  \label{c13}
\end{equation}
Now, it is easy to derive the Leibniz rule $[D_\mu,\hat{f}\cdot\hat{g}]$ and
coproduct $\bigtriangleup D_\mu$.
Similar formulae can be obtained for the action $\rhd$ of \p\ on the space \An.

We have already mentioned that \Mmn\ is in \p. It is shown in \cite{mks} that
for the {\itshape natural} realization \Mmn\ has the form
\begin{equation}
  \Mmn=X_\mu D_\nu-X_\nu D_\mu.
  \label{c15}
\end{equation}
The next step is to find the formula for $\lpl\Mmn$. Using the induction,
one can derive the formula
\begin{equation}
  [\Mmn,\hat{f}(\hat{x})]=\Mmn\rhdb\hat{f}(\hat{x})+
    ia_\mu\left(\partial_\alpha^LZ\rhdb\hat{f}(\hat{x})\right)M^\alpha_\nu-
    ia_\nu\left(\partial_\alpha^LZ\rhdb\hat{f}(\hat{x})\right)M^\alpha_\mu.
  \label{c16}
\end{equation}

\subsection{\ka-Poincar\'e Hopf algebra}
\label{kp}

Formulae (\ref{b38}) and (\ref{c13}) show that the coproduct $\lpl\Dmu$ can
be written in the form
\begin{equation}
  \lpl\Dmu=\Dmu\otimes Z^{-1}+1\otimes\Dmu+ia_\mu\partial_\alpha^L Z
    \otimes D^\alpha.
  \label{c19}
\end{equation}
and formulae (\ref{b38}) and (\ref{c16}) show that the coproduct \lpl\Mmn\ can
be written in the form
\begin{equation}
  \lpl\Mmn=\Mmn\otimes 1+1\otimes\Mmn+ia_\mu (\partial^L)^\alpha Z\otimes
    M_{\alpha\nu}-ia_\nu (\partial^L)^\alpha Z\otimes M_{\alpha\mu}.
  \label{c22}
\end{equation}

The coproduct $\lpl\Mmn$ can be calculated also by using the formula
\begin{equation}
  \Mmn=\left(\hat{x}_\mu\Dnu-\hat{x}_\nu\Dmu\right)Z
  \label{c25}
\end{equation}
(see \cite{mks}).
Let us write $\hat{f}$ instead of $\hat{f}(\hat{x})$ and
$\hat{g}$ instead of $\hat{g}(\hat{x})$. Then
\begin{eqnarray}
  m((\lpl\Mmn)\rhdb(\hat{f}\otimes\hat{g}))&=&
    \Mmn\rhdb(\hat{f}\cdot\hat{g})=\nonumber \\
  &=&\left((\hat{x}_\mu\Dnu-\hat{x}_\nu\Dmu)Z\right)\rhdb
    (\hat{f}\cdot\hat{g})=\nonumber \\
  &=&\left(\hat{x}_\mu\Dnu-\hat{x}_\nu\Dmu\right)\rhdb
    ((Z\rhdb\hat{f})\cdot (Z\rhdb\hat{g}))=\nonumber \\
  &=&\hat{x}_\mu\rhdb\Big{(}(\Dnu Z\rhdb\hat{f})\cdot \hat{g}+
    (Z\rhdb\hat{f})\cdot (\Dnu Z\rhdb\hat{g})+\nonumber \\
  &+&ia_\nu((\partial_\alpha^L Z^2\rhdb\hat{f})
    \cdot (D^{\alpha}Z\rhdb\hat{g}))\Big{)}-\nonumber \\
  &-&\hat{x}_\nu\rhdb\Big{(}(\Dmu Z\rhdb\hat{f})\cdot \hat{g}+
    (Z\rhdb\hat{f})\cdot (\Dmu Z\rhdb\hat{g})+\nonumber \\
  &+&ia_\mu((\partial_\alpha^L Z^2\rhdb\hat{f})
    \cdot (D^{\alpha}Z\rhdb\hat{g}))\Big{)}.
\end{eqnarray}
The last equality is obtained by the formula (\ref{c19}) for the coproduct of
$D_\mu$. Action of $\hat{x}_\mu$ can be understood as multiplication by
$\hat{x}_\mu$. Hence, we can either multiply the first term by $\hat{x}_\mu$
or apply formula (\ref{b52}). The principle is very simple: we want to have
$\hat{x}_\mu$ and \Dnu\ together either on the left had side or the right hand
side of the symbol $\otimes$.
It means that formula (\ref{b52}) is used for the second and the third term.
We could also develop the formula for $\lpl \hat{x}_\mu$, but authors still
work on proper mathematical setting.

Antipodes $S(\Mmn)$ and $S(D_\mu)$ are calculated from the definition of the
antipode in Hopf algebra. A short calculation produces the antipode $S(D_\mu)$
\begin{equation}
  S(\Dmu)=\left(-\Dmu+ia_\mu\partial^LD\right)Z
  \label{c28}
\end{equation}
and $S(\Mmn)$
\begin{equation}
  S(\Mmn)=-\Mmn+ia_\mu\partial_\alpha^L M_{\alpha\nu}-ia_\nu
    \partial_\alpha^LM_{\alpha\mu}.
  \label{c31}
\end{equation}

It is important to emphasize that formulae for the coproduct and antipode of
\Mmn\ and $D_\mu$ are written in the compact form i.~e. the are unified in
a covariant way. In the literature they are given separately.
Another advantage of formulae
above is that they are written for any $a$, not only for $a=(a_0,0,\ldots,0)$.
If we set $a=(a_0,0,\ldots,0)$, then relation (\ref{c28}) can be written
in the form
\begin{equation}
  S(D_0)=\left(-D_0+ia_0\partial^LD\right)Z
  \label{c34}
\end{equation}
and
\begin{equation}
  S(D_k)=-D_kZ
\end{equation}
Also, relation (\ref{c31}) can be written in the form
\begin{equation}
  S(M_{0k})=-M_{0k}+ia_0\partial_\alpha^LM_{\alpha k}
  \label{c37}
\end{equation}
and
\begin{equation}
  S(M_{mk})=-M_{mk}
  \label{c40}
\end{equation}
It is easy to compare these relations with relations for the antipode that
can be found elsewhere (see \cite{bp10}).

Finally, the counit $\epsilon$ is trivial. It means that $\epsilon(1)=1$ and
$\epsilon(\Mmn)=\epsilon(D_\mu)=0$.

Formulae for the coalgebra structure have to be checked. Namely, conditions
above are necessary but generally not sufficient.
We have checked that the formulae for the {\itshape natural} realization
satisfy all axioms of the Hopf algebra, as well as formulae written in the
Subsection \ref{ex}.

Relations (\ref{c5}) show that the algebra generated by \Mmn\ and $D_\mu$
is isomorphic to the algebra \p\ introduced before the Subsection \ref{sco}.
Hence, it produces the coalgebra structure on the algebra \p.

Let us mention that $S^2(D_\mu)=D_\mu$ and
$S^2(\Mmn)=Z^{1-n}\Mmn Z^{n-1}$.
Our shift operator behaves
very well under the coproduct and the antipode: $\lpl Z=Z\otimes Z$ and
$S(Z)=Z^{-1}$. Also, the operator $\Box$ satisfies the relation $S(\Box)=\Box$.
Our operators $\Dmu$ satisfy relations $(S\Dmu)(SD^\mu)=D^2$,
$S\left(\partial_\mu^LD^\mu Z(D)\right)=\partial^LD$ and
$Z(D)Z(S(D))=1$.

\subsection{Relations among the star product, coproduct and realization
  $\varphi_{\mu\nu}$}
\label{rel}

The main idea of this subsection is to show that the star product,
coproduct and the realization $\varphi_{\mu\nu}$ are equivalent i.~e.
one can reconstruct the remaining two vertices from any vertex of the
following triangle:\\
.
\put(194,-8){$\varphi_{\mu\nu}$}
\put(126,-120){\lpl}
\put(272,-118){$\star$}
\put(193,-15){\vector(-2,-3){60}}
\put(207,-15){\vector(2,-3){60}}
\put(139,-105){\vector(2,3){60}}
\put(140,-117){\vector(1,0){127}}
\put(274,-105){\vector(-2,3){60}}
\put(267,-113){\vector(-1,0){127}}
\put(141,-65){(i)}
\put(173,-65){(ii)}
\put(215,-65){(iii)}
\put(251,-65){(iv)}
\put(200,-105){(v)}
\put(200,-130){(vi)}

Most of arrows are already explained above. However, let us briefly recall
formulae. We should keep in mind that $\hat{x}$ and $\partial$ satisfy
relations (\ref{b1}), (\ref{c1}) and (\ref{c4}) and our realizations are,
generally, nonhermitian.\\
(i) We have already seen how to obtain the coproduct from functions
$\varphi_{\mu\nu}$ for the {\itshape natural} realization. The first step was
to find the commutator $[\pmu,\hat{f}(\hat{x})]$ (formula (\ref{c13})).
Then, it is easy to reconstruct the coproduct (\ref{c19}).
The similar procedure can be done in general. Again, the first step is to
calculate the commutator $[\pmu,\hat{f}(\hat{x})]$:
\begin{eqnarray}
  [\pmu,\hat{x}_{\nu_1}\ldots\hat{x}_{\nu_k}]&=&
  \sum_{l=1}^k\hat{x}_{\nu_1}\ldots[\pmu,\hat{x}_{\nu_l}]\ldots
  \hat{x}_{\nu_k}=\sum_{l=1}^k\hat{x}_{\nu_1}\ldots\varphi_{\mu\nu_l}
  \hat{x}_{\nu_{l+1}}\ldots\hat{x}_{\nu_k}=\nonumber\\
  &=&\sum_{l=1}^k\hat{x}_{\nu_1}\ldots\left(\hat{x}_{\nu_{l+1}}
  \varphi_{\mu\nu_l}+\frac{\partial\varphi_{\mu\nu_l}}{\partial
  \partial_\lambda}\varphi_{\lambda\nu_{l+1}}\right)\ldots\hat{x}_{\nu_k}.
\end{eqnarray}
The main idea in this process was to push $\partial$s to the right.
Hence, we have used the formula $\ds \varphi_{\mu\nu_l}\hat{x}_{\nu_{l+1}}
=\hat{x}_{\nu_{l+1}}\varphi_{\mu\nu_l}+\frac{\partial\varphi_{\mu\nu_l}}
{\partial\partial_\lambda}\varphi_{\lambda\nu_{l+1}}$ in the last step.

Using this procedure one can calculate the coproduct for numerous examples.
Let us do this calculation for the {\itshape left covariant} realization
(see Subsection \ref{ex} for details). Now our commutator $[\pmu^L,
\hat{f}(\hat{x})]$ has the nice form:
\begin{equation}
  [\pmu^L,\hat{f}(\hat{x})]=\left(\pmu^L\rhdb\hat{f}\right)Z^{-1}.
  \label{c64}
\end{equation}
This formula can be proved by induction on the length of monomials
that stand for $\hat{f}(\hat{x})$. The key step in the proof is the action
of $Z^{-1}$ on $\hat{f}(\hat{x})$: $Z^{-1}\rhdb\hat{f}(\hat{x})=Z^{-1}
\hat{f}(\hat{x})Z$. Again, it can be proved by induction. Now, the formula
(\ref{c64}) produces
\begin{equation}
  \lpl\pmu^L=\pmu^L\otimes Z^{-1}+1\otimes\pmu^L.
  \label{c67}
\end{equation}
Note: it can be proved that
\begin{equation}
  \pmu^L\rhdb\hat{x}_{\nu_1}\ldots\hat{x}_{\nu_k}=\sum_{l=1}^k
  \hat{x}_{\nu_1}\ldots\hat{x}_{\nu_{l-1}}\eta_{\mu\nu_l}Z^{-1}
  \hat{x}_{\nu_{l+1}}\ldots\hat{x}_{\nu_k}Z.
\end{equation}

The coproduct can be obtained by using approximations. The linear approximation
is calculated in Appendix \ref{la} (\ref{h10}). It is possible to continue, i.~e.
to calculate quadratic approximation, cubic approximation and so on. The result of
this procedure will be the formula for the coproduct in the form of the series.\\
(ii) The action of the coproduct of \pmu\ on $\hat{x}_{\nu}\hat{f}$ defines the
commutator $[\pmu,\hat{x}_\nu]$ and then it is easy to reconstruct
$\varphi_{\alpha\mu}$. For example, for the {\itshape natural} realization
\begin{equation}
  m\left((\lpl D_\mu)\rhdb(\hat{x}_{\nu}\otimes\hat{f})\right)=
  (\pmu\hat{x}_\nu)(Z^{-1}\hat{f})+\hat{x}_\nu(\pmu\hat{f})+
  ia_\mu(\partial_\alpha^LZ\hat{x}_\nu)(\partial^\alpha\hat{f}).
\end{equation}
It shows that
\begin{equation}
  [\pmu,\hat{x}_\nu]\hat{f}=\left(\delta_{\mu\nu}Z^{-1}+
    ia_\mu\pnu\right)\hat{f}
\end{equation}
and
\begin{equation}
  \hat{x}_\nu=x_\nu Z^{-1}+i(ax)\pnu.
\end{equation}
(iii) The realization is determined by functions $\varphi_{\alpha\mu}$. Also,
each realization determines the star product by formula (\ref{c43}).\\
(iv) Since
\begin{equation}
  x_\mu\star e^{ikx}=\left(x_\alpha\varphi_{\alpha\mu}(ik)\right)e^{ikx}
\end{equation}
and
\begin{equation}
  \left(x_\mu\star e^{ikx}\right)e^{-ikx}=x_\alpha\varphi_{\alpha\mu}(ik)
\end{equation}
one obtains
\begin{equation}
  \varphi_{\alpha\mu}(ik)=\frac{\partial}{\partial x_\alpha}\left(e^{-ikx}
    \left(x_\mu\star e^{ikx}\right)\right)
\end{equation}
(v) Function \mD\ can be obtained from the star product using formula
(\ref{c52}) and then the coproduct can be calculated using formula
(\ref{c58}).\\
(vi) The star product can be calculated directly from the formula (\ref{c46}).

\section{Realizations of \ka-Minkowski spacetime}
\label{crk}

In the previous section, the {\itshape natural} realization was the only
treated example. In this section, we will consider more general
realizations. We will start with {\itshape covariant}
realizations (see \cite{mks}) which are given by the expression
\begin{equation}
  \hat{x}_\mu=\xmu\varphi+i(ax)(\pmu\beta_1+ia_\mu\partial^2\beta_2)+
    i(x\partial)(a_\mu\gamma_1+ia^2\pmu\gamma_2)
  \label{d1}
\end{equation}
where $\varphi$, $\beta_i$ and $\gamma_i$ are functions of $A$ and $B$.
In the Section \ref{gen}, we will consider general (noncovariant) realization
given by the formula \ref{c4}.
Let us find the relationship between the {\itshape covariant} realizations
and the {\itshape natural} realization.
The most general Ansatz for $\{\pmu,\xmu\}$ is
\begin{eqnarray}
  \Dmu&=&\pmu G_1(A,B)+i a_\mu\partial^2G_2(A,B)
  \label{d2}\\
  X_\mu&=&x^\alpha\psi_{\alpha\mu}(A,B)
  \label{d3}
\end{eqnarray}
where \pmu\ and $x_\mu$ satisfy equations: $[\pmu,\pnu]=0$,
$[\xmu,\xnu]=0$ and $[\pmu,\xnu]=\eta_{\mu\nu}$. Since $[D_\mu,X_\nu]=
\eta_{\mu\nu}$,
\begin{equation}
  \psi=\left(\frac{\partial\Dmu}{\partial\pnu}\right)^{-1}
  \label{d4}
\end{equation}
i.~e. $\psi$ is determined by $G_1$ and $G_2$.

We can also consider ``inverse transformations''
\begin{eqnarray}
  \pmu&=&\Dmu \mathcal{G}_1(i(aD),a^2D^2)+ia_\mu D^2\mathcal{G}_2
    (i(aD),a^2D^2)\label{d5}\\
  \xmu&=&X^\alpha\mathcal{P}_{\alpha\mu}(i(aD),a^2D^2)\label{d6}
\end{eqnarray}
where $\mathcal{G}_i$ and $\mathcal{P}$ are real functions.

It is easy to construct the coproduct and the antipode in the
Hopf algebra related to the realization given by (\ref{d5}) and (\ref{d6}).
\begin{equation}
  \lpl\pmu=\lpl\Dmu(\mathcal{G}_1(i(a\lpl D),a^2\lpl D^2))+
    ia_\mu\lpl D^2(\mathcal{G}_2(i(a\lpl D),a^2\lpl D^2))
\end{equation}
where
\begin{eqnarray}
  \lpl D^2=\lpl\Dmu\lpl D^\mu=D^2\otimes Z^{-2}+1\otimes D^2+
   2D_\alpha\otimes D^\alpha Z^{-1}+\nonumber\\
   +i^2a^2\partial_\alpha^L\partial_\beta^L Z^2\otimes D^\alpha D^\beta+
   2i(aD)\partial_\alpha^L Z\otimes D^\alpha Z^{-1}+
   2\partial_\alpha^L Z\otimes D^\alpha(aD)
   \label{d8}
\end{eqnarray}
where $\partial_\mu^L=\Dmu-\frac{ia_\mu}2\Box$ is the {\itshape left covariant}
realization that already we had before (see the formula (\ref{d34})).

The coproduct of \Mmn, $\lpl\Mmn$, remains unchanged. Operators
$\pmu^L=\Dmu-\frac{ia_\mu}2\Box$ and $Z$ in the formula (\ref{c22})
have to be expressed in terms of $\partial$.
The same calculation can be done for the antipode.

Let us consider one example. We will calculate the coproduct \lpl\Dmu\ in  two
different ways. Formula (\ref{d34}) shows that \Dmu\ has the form
\begin{equation}
  D_\mu=\pmu^L+\frac{ia_\mu}2\Box=\pmu^L+\frac{ia_\mu}2Z(\partial^L)^2.
\end{equation}
The action of the coproduct produces (the coproduct of \pmu\ is given by the
formula (\ref{c67}))
\begin{eqnarray}
  \lpl D_\mu&=&\pmu^L\otimes Z^{-1}+1\otimes\pmu^L+\nonumber\\
  &+&\frac{ia_\mu}2(Z\otimes Z)\left(\pal^L\otimes Z^{-1}+1\otimes\pal^L\right)
  \left((\partial^L)^\alpha\otimes Z^{-1}+1\otimes(\partial^L)^\alpha\right)=
    \nonumber\\
  &=&\pmu^L\otimes Z^{-1}+1\otimes\pmu^L+\nonumber\\
  &+&\frac{ia_\mu}2\left(Z(\partial^L)^2\otimes Z^{-1}+
  2\pal^L Z\otimes(\partial^L)^\alpha+Z\otimes Z(\partial^L)^2\right)
  \label{d10}
\end{eqnarray}
The coproduct $\lpl D_\mu$ is calculated in formula (\ref{c19}). Hence
\begin{eqnarray}
  \lpl D_\mu&=&D_\mu\otimes Z^{-1}+1\otimes D_\mu+ia_\mu\pal^LZ\otimes
  D^{\alpha}\label{d13}\\
  &=&\left(\pmu^L+\frac{ia_\mu}2Z(\partial^L)^2\right)\otimes Z^{-1}+
  1\otimes\left(\pmu^L+\frac{ia_\mu}2Z(\partial^L)^2\right)+\nonumber\\
  &+&ia_\mu\pal^L Z\otimes
  \left((\partial^L)^\alpha+\frac{ia^\alpha}2Z(\partial^L)^2\right)\nonumber\\
  &=&\pmu^L\otimes Z^{-1}+1\otimes\pmu^L+\frac{ia_\mu}2Z(\partial^L)^2
  \otimes Z^{-1}+\nonumber\\
  &+&1\otimes\frac{ia_\mu}2Z(\partial^L)^2+ia_\mu\pal^LZ\otimes
  (\partial^L)^\alpha+ia_\mu\pal^L Z\otimes\frac{ia^\alpha}2Z(\partial^L)^2
  \nonumber
\end{eqnarray}
It remains to compare the results of equations (\ref{d10}) and (\ref{d13})
and use the formula $Z=1+ZA$.

In order to complete the example, let us mention that
\begin{equation}
  X_\mu=x_\mu^L-i(ax^L)\frac{\partial_\mu^L}{1-\frac{a^2}2\Box}.
\end{equation}

In the Appendix, the linear approximation in $a$ is calculated. Using the
quadratic, cubic and other approximations in $a$ one can obtain the
series when the exact formula for the coproduct is too complicated.

\subsection{Examples}
\label{ex}

We will consider two classes of realizations, called type I and type II.
For each realization, it is interesting to find out the relation to the {\itshape
natural} realization. Also, it is useful to express the shift operator $z$:
\begin{equation}
  z^{-1}(\partial)=Z^{-1}(D(\partial))=G_1\varphi
  \label{d16}
\end{equation}
where $\varphi$ appears in formula (\ref{d1}) and $G_1$ appears in formula
(\ref{d2}).

For type I realizations, $\beta_1=\beta_2=0$ (see the formula (\ref{d1}))
and $\hat{x}_\mu$ has the form
\begin{equation}
  \hat{x}_\mu=\xmu\varphi+i(x\partial)(a_\mu\gamma_1+ia^2\pmu\gamma_2)
  \label{d19}
\end{equation}
where
\begin{equation}
  \gamma_1=\frac{\left(1+\frac{\partial\varphi}{\partial A}\right)\varphi}
    {\varphi-\left(A\frac{\partial\varphi}{\partial A}+2B\frac{\partial\varphi}
    {\partial B}\right)},
  \label{d22}
\end{equation}
and
\begin{equation}
  \gamma_2=-\frac{2\frac{\partial\varphi}{\partial B}\varphi}
    {\varphi-\left(A\frac{\partial\varphi}{\partial A}+2B\frac{\partial\varphi}
    {\partial B}\right)}.
  \label{d25}
\end{equation}
The expressions for $G_i$ are given by
\begin{equation}
  G_1=\frac1{\varphi+A}
\end{equation}
and
\begin{equation}
  G_2=\frac1{2\varphi(\varphi+A)}.
  \label{d28}
\end{equation}
Operator $D_\mu$ has the form
\begin{equation}
  \Dmu=\pmu\frac{Z^{-1}}{\varphi}+\frac{ia_\mu}{2}\Box
  \label{d31}
\end{equation}
Also
\begin{equation}
  Z=1+\frac{A}{\varphi}.
\end{equation}

Three examples of type I realizations are mentioned in \cite{mks}.
For {\itshape left covariant} realization $\varphi_L=1-A$,
for {\itshape right covariant} realization $\varphi_R=1$ and
for {\itshape totally symmetric} realization $\varphi_S=\frac{A}{e^A-1}$.

Let us write down formulae for the {\itshape left covariant} realization:
\begin{eqnarray}
  &&\hat{x}_\mu=x_\mu^L(1-A)\nonumber\\
  &&\Mmn=\xmu^L\pnu^L-\xnu^L\pmu^L+\frac12i(\xmu^La_\nu-\xnu^La_\mu)
       \frac1{1-A}(\partial^L)^2\phantom{aaaaaaaaaaaaaaaa}\nonumber\\
  &&D_\mu=\pmu^L+\frac{ia_\mu}2\Box\label{d34}\\
  &&Z=\frac1{1-A}\nonumber\\
  &&\lpl\pmu^L=\pmu^L\otimes Z^{-1}+1\otimes\pmu^L
\end{eqnarray}
Similarly, for the {\itshape right covariant} realization, following formulae
are obtained
\begin{eqnarray}
  &&\hat{x}_\mu=x_\mu^R+ia_\mu(x^R\partial^R)\nonumber\\
  &&\Mmn=\xmu^R\pnu^R-\xnu^R\pmu^R+i(x^R\partial^R)
    (a_\mu\pnu^R-a_\nu\pmu^R)+\frac{i}2
    (a_\nu\xmu^R-a_\mu\xnu^R)(\partial^R)^2\nonumber\\
  &&D_\mu=\frac1{1+A}\pmu^R+\frac{ia_\mu}2\Box\nonumber\\
  &&Z=1+A\nonumber\\
  &&\lpl\pmu^R=\pmu^R\otimes1+Z\otimes\pmu^R
\end{eqnarray}
For the {\itshape totally symmetric} ({\itshape Weyl symmetric}) realization,
label $S$ on \xmu\ and \pmu\ is omitted for simplicity. Hence,
\begin{eqnarray}
  &&\hat{x}_\mu=x_\mu\frac{A}{e^A-1}+ia_\mu(x\partial)
    \frac{e^A-1-A}{(e^A-1)A}\nonumber\\
  &&\Mmn=\xmu\pnu-\xnu\pmu+i(x\partial)
    (a_\mu\pnu-a_\nu\pmu)\frac{e^A-1-A}{A^2}\nonumber\\
  &&D_\mu=\frac{e^A-1}{Ae^A}\pmu+\frac{ia_\mu}2\Box
    \phantom{aaaaaaaaaaaaaaaaaaaaaaaaaaaaaaaaaaaaaaaa}\nonumber\\
  &&Z=e^A\nonumber\\
  &&\lpl\pmu=\varphi_S(\lpl A)\left(\frac{\pmu}{\varphi_S(A)}\otimes1\right)+
      \varphi_S(-\lpl A)\left(1\otimes\frac{\pmu}{\varphi_S(-A)}\right).
\end{eqnarray}

For type II realizations, $\beta_1=1$ and $\beta_2=0$ in formula (\ref{d1}).
Hence, $\hat{x}_\mu$ has the form
\begin{equation}
  \hat{x}_\mu=\xmu\varphi+i(ax)\pmu+
    i(x\partial)(a_\mu\gamma_1+ia^2\pmu\gamma_2),
  \label{d35}
\end{equation}
where $\gamma_1$ has the same form as for type I realizations (formula
(\ref{d22})) and $\gamma_2$ has the form
\begin{equation}
  \gamma_2=\frac{\frac{\partial\varphi}{\partial A}-2(\varphi+A)
    \frac{\partial\varphi}{\partial B}}
    {\varphi-\left(A\frac{\partial\varphi}{\partial A}+2B\frac{\partial\varphi}
    {\partial B}\right)}.
  \label{d36}
\end{equation}
It remains to write down expressions for $G_i$ and $Z$.
\begin{equation}
  G_1=\frac1{\sqrt{(\varphi+A)^2+B}},
\end{equation}
$G_2=0$ and
\begin{equation}
  Z=\sqrt{\left(1+\frac A\varphi\right)^2+\frac B{\varphi^2}}.
\end{equation}

The {\itshape natural} realization is of type II. In this case, $\varphi_N=
-i(aD)+\sqrt{1-a^2D^2}$.

There is another interesting example of type II realizations:
Magueijo--Smolin \cite{Magueijo:2001cr}. In this case $\varphi_{MS}=1$. We
omit label $MS$ on \xmu\ and \pmu\ for simplicity. Hence,
\begin{eqnarray}
  &&\hat{x}_\mu=x_\mu+i(xa)\pmu+i(x\partial)a_\mu
    \phantom{aaaaaaaaaaaaaaaaaaaaaaaaaaaaaaaaaa}\nonumber\\
  &&\Mmn=\xmu\pnu-\xnu\pmu+i(x\partial)(a_\mu\pnu-a_\nu\pmu)\nonumber\\
  &&Z=\sqrt{(1+A)^2+B}\nonumber\\
  &&D_\mu=Z^{-1}\pmu\nonumber\\
  &&\lpl\pmu=\pmu\otimes1+Z\otimes\pmu+ia_\mu\left(\pmu-\frac{ia_\alpha}2
    \Box\right)Z\otimes\partial^\alpha Z
\end{eqnarray}
The Magueijo--Smolin realization does not satisfy condition (\ref{b27}) i.~e.
\pmu\ does not transform as a vector under the action of \Mmn.

It is easy to calculate antipode in this special realizations.

\subsection{General Dirac derivatives}
\label{gdd}

In general realizations do not satisfy condition (\ref{b27}). If this condition is
added, then $D_\mu$ and $X_\mu$ have the simpler form
\begin{eqnarray}
  \Dmu&=&\pmu G(B) \nonumber\\
  X_\mu&=&x_\alpha\psi_{\alpha\mu}(B).
  \label{d39}
\end{eqnarray}
$G(B)$ has the form
\begin{equation}
  G(B)=\frac1{\sqrt{f^2(B)+B}}
  \label{d40}
\end{equation}
and
\begin{equation}
  z^{-1}(\partial)\equiv Z^{-1}(D(\partial))=\frac{-A+f(B)}{\sqrt{f^2(B)+B}}
  \label{d43}
\end{equation}
for a real function $f(B)$.
Then $\hat{x}_\mu$ has the form
\begin{equation}
  \hat{x}_\mu=\xmu(-A+f(B))+i(xa)\pmu-a^2(x\partial)\pmu\gamma_2(B)
  \label{d46}
\end{equation}
The inverse transformation has a simpler form
\begin{equation}
  \pmu=\Dmu\mathcal{G}(a^2D^2)
\end{equation}
One obtains $\ds \Dmu=\frac{\pmu}{\sqrt{f^2+B}}$, $\ds a^2D^2=
\frac{B}{f^2+B}$. 

Operator \Mmn\ can be expressed by formula (\ref{c25}):
\begin{equation}
  \Mmn=\left(\hat{x}_\mu\Dnu-\hat{x}_\nu\Dmu\right)Z.
\end{equation}
Using formulae (\ref{d46}) for $\hat{x}_\mu$, (\ref{d39}) for \Dmu\ (and
(\ref{d40}) for $G$) and (\ref{d43}) for $Z$, it is easy to check the formula
for \Mmn
\begin{equation}
  M_{\mu\nu}=x_\mu\pnu-x_\nu\pmu.
\end{equation}

\subsection{Construction of general realization $\varphi_{\mu\nu}$}
\label{gen}

Let us consider, again, (general, not necessarily {\itshape covariant}) realizations
given by formulae (\ref{c2}) and (\ref{c4}). We want to find the relationship
among them. In order to distinguish them, one of them will
be the {\itshape natural} realization. Realizations are related by the similarity
transformations i.~e.
\begin{eqnarray}
  D_\mu&=&\mE\pmu\mE^{-1}\nonumber\\
  X_\mu&=&\mE x_\mu\mE^{-1}
  \label{d55}
\end{eqnarray}
where $\mE=\exp\{x^\alpha\Sigma_\alpha(\partial)\}$ and
functions $\ds \Sigma_\alpha(\partial)$ satisfy the boundary condition
\begin{equation}
  \lim_{a\rightarrow0}\Sigma_\alpha=0.
\end{equation}

It is easy to see that relations (\ref{d55}) transform to
\begin{eqnarray}
  D_\mu&=&D_\mu(\partial)\nonumber\\
  X_\mu&=&x^\alpha\psi_{\alpha\mu}(\partial).
  \label{d49}
\end{eqnarray}
Since, $[D_\mu,\hat{x}_\nu]=\Phi_{\mu\nu}(D)$,
\begin{equation}
  \frac{\partial D_\mu}{\partial\partial_\alpha}\varphi_{\alpha\nu}=
  \Phi_{\mu\nu}(D(\partial))
\end{equation}
and
\begin{equation}
  \varphi_{\alpha\nu}=\left[\frac{\partial D}{\partial\partial}
  \right]_{\alpha\mu}^{-1}\Phi_{\mu\nu}(D(\partial)).
\end{equation}
It shows that using similarity transformation (or equivalently functions
$D_\mu$ and $\psi_{\mu\nu}=\left(\frac{\partial D}
{\partial\partial}\right)_{\mu\nu}^{-1}$) and the set of functions
$\Phi_{\mu\nu}$, one can construct the set of functions $\varphi_{\mu\nu}$.
Since functions $\varphi_{\mu\nu}$ and $h_{\mu\nu}$ are related by
the formula (\ref{c5}), we find functions $h_{\mu\nu}(p)$.
Hence, the above constructions produces the set of all solutions $\{h_{\mu\nu}\}$
of the system of differential equations (\ref{b20}), (\ref{b22}) and (\ref{b25}).

If $D_\mu(\partial)$ is given and $[\psi_{\mu\nu}]$ is a regular matrix,
then $\Sigma_\alpha(\partial)$ exist and can be generally  expressed in
terms of the series of $\partial$, and vice verse.

\section{L--R duality}
\label{lr}

Let us consider the following problem: if the realization $\varphi$ is given
by relation (\ref{c4}), is it possible to find the dual realization,
$\tilde{\varphi}$, such that the following relation is satisfied
\begin{equation}
  (f\star_\varphi g)(x)=(g\star_{\tilde{\varphi}}f)(x)?
  \label{e1}
\end{equation}
The answer is positive and it follows from the previous subsection.

It is shown in the previous section that to each realization there
corresponds a star product and the coproduct. We will relate two realizations
given in relation (\ref{e1}) by the coproduct.
The star product of the left hand side of the relation (\ref{e1}) is related
to the coproduct \lpl\ by the formula (\ref{c46}):
\begin{equation}
  (f\star_\varphi g)(x)=m_0(e^{x_\alpha(\lpl-\lpl_0)\partial^\alpha}(f\otimes g))(x).
\end{equation}
Let us calculate the coproduct $\tilde{\lpl}$ of the dual realization. We set
\begin{equation}
  \tilde{\lpl}=\tau_0\lpl
\end{equation}
where $\tau_0$ is the flip operator ($\tau_0(a\otimes b)=b\otimes a$). The
corresponding star product is calculated by
\begin{equation}
  (g\star_{\tilde{\varphi}} f)(x)=m_0(e^{x_\alpha(\tilde{\lpl}-\lpl_0)\partial^\alpha}
  (g\otimes f))(x)
\end{equation}
and the dual realization $\tilde{\varphi}$ can be calculated by the procedure
(2) in the Subsection \ref{rel}.

Let us consider operators $\hat{y}_\mu$ defined by
\begin{equation}
  \hat{y}_\mu=x^\alpha\tilde{\varphi}_{\alpha\mu}(\partial).
\end{equation}
Then the following two relations are satisfied:
\begin{equation}
  \hat{x}_\mu\rhd f(x)=x_\mu\star_\varphi f(x)=f(x)\star_{\tilde{\varphi}}x_\mu
  \label{e4}
\end{equation}
and
\begin{equation}
  \hat{y}_\mu\rhd f(x)=x_\mu\star_{\tilde{\varphi}} f(x)=f(x)\star_\varphi x_\mu.
  \label{e7}
\end{equation}
Using the above two relations, it is easy to prove that operators
$\hat{y}_\mu$ satisfy dual relations to the set of relations (\ref{b1}):
\begin{equation}
  [\hat{y}_\mu,\hat{y}_\nu]=-i(a_\mu\hat{y}_\nu-a_\nu\hat{y}_\mu).
  \label{e10}
\end{equation}
One calculates the left hand side of the relation (\ref{e10})
\begin{equation}
  [\hat{y}_\mu,\hat{y}_\nu]\rhd f(x)=(f(x)\star_\varphi x_\nu)
  \star_\varphi x_\mu-(f(x)\star_\varphi x_\mu)
  \star_\varphi x_\nu.
\end{equation}
Since the star product is associative, it transforms to
\begin{equation}
  [\hat{y}_\mu,\hat{y}_\nu]\rhd f(x)=
  f(x)\star_\varphi i(a_\nu x_\mu-a_\mu x_\nu)=
  -i(a_\mu\hat{y}_\nu-a_\nu\hat{y}_\mu)\rhd f(x).
\end{equation}
It proves the relation (\ref{e10}).

Operator $\hat{y}_\mu$ commutes with operator $\hat{x}_\nu$, i.~e.
\begin{equation}
  [\hat{y}_\mu,\hat{x}_\nu]=0.
\end{equation}
This proof is trivial.

Finally, let us write down dual operators for the {\itshape natural}
realization. Operators $\hat{y}_\mu$ have the form
\begin{equation}
  \hat{y}_\mu=X_\mu-ia_\mu(XD)+i(aX)\partial_\mu^LZ
\end{equation}
The {\itshape left covariant} realization is dual to the
{\itshape right covariant} realization (see Subsection \ref{ex}).
Finally, the {\itshape totally symmetric} realization is symmetric to itself
where $a_\mu$ goes to $-a_\mu$.

In general, operators $\hat{y}_\mu$ have the form
\begin{equation}
  \hat{y}_\mu=\left(\hat{x}_\mu-ia_\mu(\hat{x}\partial^L)\right)Z.
\end{equation}
Finally, let us mention the relation that connects $\tilde{\lpl}$ and the
antipode $S$:
\begin{equation}
  \tilde{\lpl}=(S\otimes S)\circ\lpl\circ S^{-1}
\end{equation}

\section{Integral formulae}
\label{if}

The {\itshape generalized involution}, $f^*$, of the function $f$ is defined by
\begin{equation}
  \left(e^{ikx}\right)^*=e^{iS(k)x}
\end{equation}
for the exponential function $f(x)=e^{i(kx)}$ and by
\begin{equation}
  f^*(x)=\frac1{(2\pi)^{\frac n2}}\int\mathrm{d}^nk\overline{\left(\tilde{f}(k)
  \right)}e^{iS(k)x}
  \label{f4}
\end{equation}
for any function $f$ where $\overline{\tilde{f}(k)}$ denotes the complex conjugation.
Here, $f$ has the form
\begin{equation}
  f(x)=\frac1{(2\pi)^{\frac n2}}\int\mathrm{d}^nk\left(\tilde{f}(k)\right)e^{ikx}.
  \label{f7}
\end{equation}
Let us mention that $f^*$ can be also defined as
\begin{equation}
  1\lhd\left(\hat{f}(\hat{x})\right)^\dagger=f^*(x)
\end{equation}
where $\dagger$ denotes the Hermitian conjugation ($x_\mu^\dagger=x_\mu$
and $\partial_\mu^\dagger=-\partial_\mu$).
Also, $\ds \lambda^*=\overline{\lambda}$ for $\lambda\in\C$ and
\begin{equation}
  \lim_{a\rightarrow0}f^*=\overline{f}.
\end{equation}
Finally, $f^*=\overline{f}$ for all selfdual realizations.
Hence, $f^*=\overline{f}$ for the {\itshape totally symmetric} realization.

Let us remark the following property of the star product:
\begin{equation}
  (f\star g)^*=g^*\star f^*.
  \label{f17}
\end{equation}
We will sketch the proof for functions $f(x)=e^{ikx}$ and $g(x)=e^{iqx}$. Using
the Fourier expansion, it is easy to do the general case.
The formula (\ref{c52}) shows that $(f\star g)^*$ has the form
$e^{iS(\mD(k,q))x}$, formula (\ref{c73}) transforms it to
$e^{i\mD(S(q),S(k))x}$ and then formula (\ref{c52}) produces the right
hand side of formula (\ref{f17}).

The {\itshape star} inner product $(\cdot,\cdot)_\star$ of two
functions $f$ and $g$ is defined by
\begin{equation}
  (f,g)_\star=\int\mathrm{d}^nx\:f^*(x)\star g(x).
  \label{f1}
\end{equation}
Using the identity
\begin{equation}
  \int\mathrm{d}^nx\:e^{iS(k)x}\star e^{iqx}=
  (2\pi)^n\delta^{(n)}(\mD(S(k),q))=
  \frac{(2\pi)^n}{\det\left|\frac{\partial\mD_\mu(S(k),q)}{\partial q_\nu}
    \right|_{k=q}}\delta^{(n)}(k-q),
  \label{f2}
\end{equation}
it is easy to see that the {\itshape star} inner product satisfies all
properties of the inner product.

The denominator $\ds \det\left|\frac{\partial\mD_\mu(S(k),q)}{\partial q_\nu}
\right|_{k=q}$ can be calculated for each realization. For example, for the
{\itshape natural} realization, $\ds \det\left|\frac{\partial\mD_\mu(S(k),q)}
{\partial q_\nu}\right|_{k=q}=\frac1{\sqrt{1+a^2k^2}}$ (see \cite{msa}). Also,
it is easy to calculate that for the {\itshape totally symmetric} realization
$\ds \det\left|\frac{\partial\mD_\mu(S(k),q)}{\partial q_\nu}\right|_{k=q}=
\left(\frac{e^{aq}-1}{aq}\right)^{n-1}e^{aq}$.

This inner product will be very important when the realization (\ref{c4}) is
Hermitian since the following formula is valid:
\begin{equation}
  \int\mathrm{d}^nx\:f^*(x)\star g(x)=
  \int\mathrm{d}^nx\:\overline{f}(x)\cdot g(x).
  \label{f5}
\end{equation}
In \cite{Dimitrijevic:2003wv,Moller:2004sk,Agostini:2006zza},
the auxiliary measure $\mathrm{d}^nx\mu(x)$ was introduce in order to obtain
the trace property. Our approach is more natural.

Let us consider some properties of the {\itshape star} inner product. 
We start with $\int\mathrm{d}^nx\:f(x)\star\pmu g(x)$.
\begin{equation}
  \int\mathrm{d}^nx\:f(x)\star\pmu g(x)=
  \frac1{(2\pi)^n}\int\mathrm{d}^nk\:\int\mathrm{d}^nq\:
  \int\mathrm{d}^nx\:\tilde{f}(k)\tilde{g}(q)e^{ikx}\star\pmu e^{iqx}=
  \nonumber
\end{equation}
\begin{equation}
  =\frac1{(2\pi)^n}\int\mathrm{d}^nk\:\int\mathrm{d}^nq\:
  \int\mathrm{d}^nx\:\tilde{f}(k)\tilde{g}(q)(iq_\mu)e^{i\mD(k,q)x}=
  \nonumber
\end{equation}
\begin{equation}
  =\int\mathrm{d}^nk\:\int\mathrm{d}^nq\:
  \tilde{f}(k)\tilde{g}(q)(iq_\mu)\delta^{(n)}(\mD(k,q))=
  \nonumber
\end{equation}
\begin{equation}
  =\int\mathrm{d}^nk\:\int\mathrm{d}^nq\:
  \tilde{f}(k)\tilde{g}(q)(iS(k_\mu))\delta^{(n)}(\mD(k,q))=\ldots
  \nonumber
\end{equation}
\begin{equation}
  \ldots=\int\mathrm{d}^nx\:S(\pmu)f(x)\star g(x)
\end{equation}
We have used that $\mD(k,S(k))=0$ and $\delta^{(n)}(\mD(k,q))\neq0$ for
$q=S(k)$. Hence,
\begin{equation}
  \int\mathrm{d}^nx\:f(x)\star\pmu g(x)=\int\mathrm{d}^nx\:S(\pmu)f(x)\star g(x).
  \label{f13}
\end{equation}
Similarly,
\begin{equation}
  \int\mathrm{d}^nx\:\pmu f(x)\star g(x)=\int\mathrm{d}^nx\:f(x)\star
    S(\pmu)g(x).
  \label{f14}
\end{equation}
Formula (\ref{f14}) follows from formula (\ref{f13}) directly by $S^2(\pmu)=
\pmu$.
Since $Z=Z(\partial)$, the formula (\ref{f13}) generalizes to
\begin{equation}
  \int\mathrm{d}^nx\:f(x)\star Z g(x)=
  \int\mathrm{d}^nx\:Z^{-1}f(x)\star g(x).
\end{equation}
Finally,
\begin{equation}
  \int\mathrm{d}^nx\:f(x)\star g(x)=
  \int\mathrm{d}^nx\:Z^{n-1}g(x)\star f(x)=
  \int\mathrm{d}^nx\:Z^{n-2}g(x)\star Z^{-1}f(x)=\ldots\nonumber
\end{equation}
\begin{equation}
  \ldots=\int\mathrm{d}^nx\:g(x)\star Z^{-n+1}f(x).
  \label{f20}
\end{equation}
The formula above is proved in \cite{msa} for the Hermitian {\itshape natural}
realization. It can be also found in \cite{ds}, for the case when $n=2$.
However, it is valid for any (not necessarily Hermitian) realization.

\section{Translation invariance of the star product}
\label{ti}

One can ask if the star product is translation invariant, i.~e.
\begin{equation}
  \mathcal{T}_v(f)\star\mathcal{T}_v(g)=\mathcal{T}_v(f\star g)
  \label{g1}
\end{equation}
where $\mathcal{T}_v$ denotes the translation operator: $\mathcal{T}_v(f)(x)=
f(x+v)$ for vector $v=(v_0,\ldots,v_{n-1})\in\R^n$.
It is important to note that the translation operator $\mathcal{T}_v$ is
the similarity transformation since
\begin{equation}
  \mathcal{T}_vx_\mu=x_\mu+v_\mu=e^{v_\alpha\partial^\alpha}x_\mu
    e^{-v_\alpha\partial^\alpha}=T_vx_\mu T_v^{-1}.
\end{equation}
Similarly,
\begin{equation}
  \mathcal{T}_v(\hat{x}_\mu)=T_v\hat{x}_\mu T_v^{-1}=\hat{x}_\mu+
    v_\alpha\varphi_{\alpha\mu}
\end{equation}
Now,
\begin{eqnarray}
  \mathcal{T}_v\left(f(x)\star g(x)\right)&=&\mathcal{T}_v\left(\hat{f}
    (\hat{x})\hat{g}(\hat{x})\rhd1\right)=T_v\left(\hat{f}(\hat{x})\hat{g}
    (\hat{x})\rhd1\right)T_v^{-1}\rhd1=\nonumber \\
    &=&T_v\hat{f}(\hat{x})T_v^{-1}T_v\hat{g}(\hat{x})T_v^{-1}\rhd1=
    \mathcal{T}_v(f)\star\mathcal{T}_v(g)
\end{eqnarray}
It shows that the star product is translation invariant.

Our definition of translation invariance is different from others in literature
(see \cite{klm,dlw0701,l10}).

\section{Conclusion and outlook}
\label{con}

Several details should be considered again. Our approach to the Hopf algebra
was based on the action of \Hh\ on \Ah\ and this action was
determined by the set of functions $h_{\mu\nu}$ (the set of functions
$g_{\mu\nu\lambda}$ was determined by functions $h_{\mu\nu}$). Functions
$h_{\mu\nu}$ were determined by the set of differential equations
(\ref{b20}), (\ref{b22}), (\ref{b25}) and these equations were determined
by Jacobi identities. These Jacobi identities were key point in our
construction. They gave a nice blend of the \ka-Minkowski space and the
\ka-Poincar\'{e} algebra together with the action \rhdb. It is important to
emphasize that from one set of functions $h_{\mu\nu}$ one can construct
all possible sets of functions $h_{\mu\nu}$ (see Subsection \ref{gen}).

We point out that for the given choice of $h_{\mu\nu}$, there is
a unique expression for $g_{\mu\nu\lambda}$ (\ref{b16})
such that the Lorentz algebra is undeformed and relations (\ref{b1}-\ref{b7})
are satisfied. Also, the Hopf
algebra structure is fixed and the corresponding \ka-Poincar\'{e} algebra is fixed.
The reverse statement is not true.
Examples with undeformed Poincar\'{e} algebras are given by a
family of $h_{\mu\nu}$, see Subsection \ref{gdd}.

Concrete calculations are done using realizations. Several examples are given,
{\itshape natural} and {\itshape left covariant} realizations are calculated
completely.

Our realizations are polarized, i.~e. expressions for $\hat{x}_\mu$ contain
$x$ to the first power on the left hand side and $\partial$s on the right hand
side. We will consider the hermitian realizations, which
are not polarized, in the next paper.

The formulae obtained for the coalgebraic sector of the \ka-Poincar\'{e} algebra
are written in the covariant way, i.~e. there is one expression for the
coproduct or the antipode contrary to formulas in literature (for example,
\cite{bp10}). Also, formulas in the literature are given for $a=(a_0,0,\ldots,0)$.

To each realization, there corresponds a unique Hopf algebra structure
(coproduct) and a star product. Their relationship is analyzed in
Subsection \ref{rel}. It is shown that if one piece of information is known
(the realization, the coproduct or the the star product), then it is easy to
construct remaining two. It will be shown in future paper, that it is possible
to treat the twist on the same level, i.~e. one can construct the twist from
above pieces of data and vice verse.

Formulae (\ref{c13}) and
(\ref{c16}) led us to the Leibniz rule and the coproduct of $D_\mu$ and \Mmn.
Similarly, the formula (\ref{b52}) produces the Leibniz rule for
the element $\hat{x}_\mu$. At this moment, we can not discuss the coproduct
$\lpl\hat{x}_\mu$ since there is no Hopf algebra which contains
$\hat{x}_\mu$. However it is possible to write down the expression for
$\lpl\hat{x}_\mu$ which fits into calculations. It leads to the construction of
the generalized Hopf algebra which will appear in the future publications.

Dispersion relation is determined by $\Box$. For the natural realization, $\Box$
is determined by relation (\ref{c9}). Relation (\ref{d55}) shows how to relate
$D_\mu$ and $\partial_\mu$ for the general realization and hence write the
dispersion relation.
Dispersion relations were analyzed and discussed in \cite{bgmp09} and
references therein.

The notion of duality relates two \ka-Minkowski algebras.
Also, it relates corresponding realizations.
Hence the understanding of one realization (and appropriate formulae) gives
the understanding of the dual realization.

Let us stress the importance of the formula (\ref{f5}). In
\cite{Dimitrijevic:2003wv,Moller:2004sk,Agostini:2006zza},
the artificial measure was introduced in order to obtain desired properties.
The formula (\ref{f5}) shows another approach which is more natural.
Hermitian realizations will be considered in our following paper.
The formula (\ref{f20}) shows that the deformed trace property is satisied.
Integral formulas will be generalized further to differential forms.

The notion of translations is introduced in a different way than it is
done in \cite{klm,dlw0701,l10}. Under this approach, the star
product becomes translation invariant.

Our parameters $a_\mu$ were treated as constants related to the choice of
the \ka-Minkowski space.
Since the set of relations $[\Mmn,a_\lambda]=0$ leads to $[\Mmn,Z]\neq0$,
the Lorentz invariance is violated. Also, different sets of functions $\{h_{\mu\nu}\}$
migh lead to different physical consequences.
For example, constraints on the quantum gravity scale from \ka-Minkowski
spacetime are studied in \cite{bgmp09} and electrodynamics on
\ka-Minkowski spacetime was analyzed in \cite{hjm11}.
An alternative way is to treat $a$ as a
vector invariant under Lorentz transformations i.~e.
\begin{equation}
  [\Mmn,a_\lambda]=a_\mu\eta_{\nu\lambda}-a_\nu\eta_{\mu\lambda},
\end{equation}
together with remaining two sets of commutators: $[\hat{x}_\mu,a_\nu]=
[a_\mu,a_\nu]=0$. This construction leads to the primitive coproduct of \Mmn
(see \cite{msa,dgp09}).

\section*{Acknowledgements}

We are very thankful to Andrzej Borowiec for the numerous comments on the text.
Also, discussion with Jerzy Lukierski was very valuable.
Finally, we thank An\dj elo Samsarov for helpful remarks.
This work was supported by the Ministry of Science and Technology of the
Republic of Croatia under contract No. 098-0000000-2865.

\appendix
\numberwithin{equation}{section}

\section{Linear approximation in $a$}
\label{la}

The realization (\ref{c4}) is determined by functions $\varphi_{\mu\nu}$ and
these functions can be written as an infinite power series in $a$. We are
going to consider any function $\varphi_{\mu\nu}$ and then restrict our
attention to $0^{\rm th}$ and $1^{\rm st}$ power in $a$. The most general linear
form of the realization (\ref{c4}) has the form
\begin{equation}
  \hat{x}_\mu=x_\mu+i\left(\alpha x_\mu(a\partial)+\beta(ax)\partial_\mu+
  \gamma a_\mu(x\partial)\right)
  \label{h1}
\end{equation}
Since variables $\hat{x}_\mu$ satisfy relation (\ref{b1}), $\gamma-\alpha=1$.

We will start with the function $\mP=\mP(k,q)$ defined by
\begin{equation}
  e^{ik\hat{x}}\rhd e^{iqx}=e^{i(\mP(k,q)x)}
  \label{h4}
\end{equation}
Simple calculation shows that
\begin{eqnarray}
  \mP_\mu(k,q)=k_\mu+q_\mu&-&\frac12\left\{\alpha k_\mu(ak)+\beta a_\mu k^2+
  \gamma(ak)k_\mu\right\}\nonumber\\
                        &-&\left\{\alpha k_\mu(aq)+\beta a_\mu (kq)+
  \gamma(ak)q_\mu\right\}
  \label{h7}
\end{eqnarray}
Since $\mK(k)=\mP(k,0)$,
\begin{equation}
  \mK_\mu(k)=k_\mu-\frac12\left\{\alpha k_\mu(ak)+\beta a_\mu k^2+\gamma(ak)
  k_\mu\right\}
\end{equation}
and
\begin{equation}
  \mK_\mu^{-1}(k)=k_\mu+\frac12\left\{\alpha k_\mu(ak)+\beta a_\mu k^2+
  \gamma(ak)k_\mu\right\}.
\end{equation}
Since $\mD(k,q)=\mP(\mK^{-1}(k),q)$,
\begin{equation}
  \mD_\mu(k,q)=k_\mu+q_\mu-\left\{\alpha k_\mu(aq)+\beta a_\mu(kq)+
  \gamma(ak)q_\mu\right\}.
\end{equation}
Formula (\ref{c58}) produces
\begin{equation}
  \lpl\partial_\mu=\partial_\mu\otimes1+1\otimes\partial_\mu+i\left\{
  \alpha\partial_\mu\otimes(a\partial)+\beta a_\mu\partial^\alpha\otimes
  \partial_\alpha+\gamma(a\partial)\otimes\partial_\mu\right\}
  \label{h10}
\end{equation}

For each realization, it is easy to find $\alpha$, $\beta$ and $\gamma$.
Now, formulae above can serve as a nice check. For example, for the
{\itshape natural} realization, $\alpha=-1$, $\beta=1$ and $\gamma=0$. Then
formula (\ref{h10}) shows that
\begin{equation}
  \lpl \Dmu=\Dmu\otimes1+1\otimes\Dmu+i\left\{-\Dmu\otimes(a D)+
  a_\mu D^\alpha\otimes D_\alpha\right\}
\end{equation}
The same formula can be obtained from formula (\ref{c19}).

Let us consider the relationship between the {\itshape natural} realization
and any {\itshape covariant} realization (equations (\ref{d2}) and (\ref{d3})).
The linear approximation of these equations have the form
\begin{eqnarray}
  \Dmu=\partial_\mu+i\left\{u(a\partial)\partial_\mu+va_\mu\partial^2\right\}\\
  X_\mu=x_\mu-i\left\{ux_\mu(a\partial)+ua_\mu(x\partial)+2v(ax)\pmu\right\}
\end{eqnarray}
The relationship is given by $\alpha=-u-1$, $\beta=-2v+1$ and $\gamma=-u$.

The correspondence between $\hat{f}(\hat{x})$ and $f(x)$ is given by
\begin{equation}
  \hat{f}(\hat{x})=f(x)+i\left(\alpha\left(x\frac{\partial f}{\partial x}
  \right)(a\partial)+\beta(ax)\left(\frac{\partial f}{\partial x}\partial
  \right)+\gamma\left(a\frac{\partial f}{\partial x}\right)(x\partial)\right)
\end{equation}
and the correspondence between $f(x)$ and $\hat{f}(\hat{x})$ is given by
\begin{equation}
  f(x)=\hat{f}(\hat{x})\rhd1=\hat{f}(x)+\sum_{i<j}O_i\hat{f}(x)
  [\hat{\delta}_i,x_j]
\end{equation}
where $[\hat{\delta}_i,x_j]=\frac{\partial\hat{\delta}_i}{\partial\partial_j}$.

Let us mention that the similar calculation could be done for the quadratic
approximation in $a$.

\end{document}